\documentclass[a4paper,11pt]{article}
\pdfoutput=1 

\usepackage{jheppub} 

\usepackage[T1]{fontenc} 
\usepackage{hyperref}
\usepackage{subfig}
\usepackage{amsmath}
\allowdisplaybreaks

\title{\boldmath One-loop Feynman integrals for $2\to3$ scattering involving many scales including internal masses}

\author[a,b]{Nikolaos Syrrakos}

\affiliation[a]{Institute of Nuclear and Particle Physics, NCSR Demokritos, \\Patr. Grigoriou E' \& 27 Neapoleos Str, 15341 Agia Paraskevi, Greece}
\affiliation[b]{Physics Division, National Technical University of Athens, \\Zografou Campus, Athens 15780, Greece}

\emailAdd{syrrakos@inp.demokritos.gr}

\abstract{ We study several multiscale one-loop five-point families of Feynman integrals. More specifically, we employ the Simplified Differential Equations approach to obtain results in terms of Goncharov polylogarithms of up to transcendental weight four for families with two and three massive external legs and massless propagators, as well as with one massive internal line and up to two massive external legs. This is the first time this computational approach is applied to cases involving internal masses.}


\begin{document} 
\maketitle
\flushbottom

\section{Introduction}
\label{sec:intro}

In recent  years the field of precision calculations in collider physics has emerged as a vibrant and fruitful line of research in our attempt to understand Nature at its most fundamental level \cite{Heinrich:2020ybq}. The basic principle of this research endeavour is to have very precise experimental measurements of cross sections for Standard Model scattering processes compared against theoretical predictions of equally high precision and search for any deviations between them. Should any such deviations be established, their analysis and physical explanation would require New Physics, giving us an idea of what lies beyond the Standard Model. 

The ever increasing demand for highly precise theoretical predictions for scattering processes relevant to LHC searches poses a challenge in our ability to perform higher order calculations in perturbative Quantum Field Theory \cite{Amoroso:2020lgh}. Multiloop scattering amplitudes play a fundamental role in such calculations, encoding within their mathematical structure key information concerning the nature of particle interactions. One major aspect of the calculation of multiloop scattering amplitudes is the calculation of the relevant Feynman diagrams that are involved, which can be associated through the corresponding Feynman rules to the so-called Feynman integrals. In the following we will use the notion of \textit{diagrams} and \textit{integrals} interchangeably. 

The standard approach for the calculation of these integrals involves obtaining a complete set of Master integrals through the use of Integration-By-Part identities \cite{Chetyrkin:1981qh}, constructing a pure basis of Master integrals \cite{Henn:2014qga} and then deriving and solving differential equations \cite{de1,de2,de3,de4} in canonical form \cite{Henn:2013pwa}. This approach has yielded numerous results \cite{Kotikov:2021tai}, in part due to the fact that we have a solid understanding of the special class of functions, known as multiple or Goncharov polylogarithms \cite{Goncharov:1998kja,Duhr:2011zq,Duhr:2012fh,Duhr:2014woa}, in terms of which many Feynman integrals can be expressed. In more complicated cases however, this class of functions is not enough and important steps have been made in getting a better understanding of a more general class of functions, Elliptic integrals \cite{Remiddi:2017har, Broedel:2017kkb,Broedel:2017siw,Broedel:2018iwv,Broedel:2018qkq,Broedel:2019hyg,Duhr:2019rrs}, which appear in solutions of multiloop Feynman integrals with many scales, especially when several internal masses are introduced. 

When considering multiloop Feynman integrals involving many external particles, the current frontier lies at two-loop five-point integrals with up to one off-shell leg and massless internal lines. For the fully massless case, all Master integrals are by now known up to transcendental weight four \cite{Papadopoulos:2015jft,Gehrmann:2015bfy,Chicherin:2017dob,Gehrmann:2018yef,Chicherin:2018mue,Abreu:2018rcw,Chicherin:2018old} and their solutions have been implemented in a fast \texttt{C++} library known as \textit{pentagon functions} \cite{Chicherin:2020oor}. When one of the external particles is considered off-shell, the planar topologies have been recently solved using two different computational approaches for the solution of canonical differential equations, numerically \cite{Abreu:2020jxa} and analytically \cite{Canko:2020ylt}. The numerical calculation was performed using a generalised power-series method \cite{Francesco:2019yqt,Hidding:2020ytt}, while the analytical solution was achieved through the use of the Simplified Differential Equations approach \cite{Papadopoulos:2014lla}, with the results given in terms of Goncharov polylogarithms of up to transcendental weight four. These results are relevant to many $2\to3$ scattering processes studied experimentally at the LHC, e.g. $W+2$ jets production. For the computation of the relevant scattering amplitudes, one-loop five-point Feynman integrals with one off-shell leg also have to be known up to transcendental weight four \cite{Syrrakos:2020kba}. These results were recently used for the calculation of two-loop QCD corrections to $Wb\bar{b}$ production \cite{Badger:2021nhg}. First results for one of the non-planar topologies have also appeared using a numerical approach \cite{Papadopoulos:2019iam}. More recently the three \textit{hexabox} topologies were calculated in \cite{Abreu:2021smk} using the same approach as in \cite{Abreu:2020jxa}.

While staying at the level of five-point Feynman integrals, at some point we will have to introduce internal masses and consider more than one massive external particle. Judging from the level of complexity of the so far accumulated results, these Feynman integrals are expected to be highly non-trivial to be solved using current approaches, especially the genuine two-loop ones. To that end, we believe that it is instructive to consider first the relevant one-loop five-point Feynman integrals with more than one off-shell leg and/or with internal masses. The interest of these Feynman integrals is twofold. From a more formal point of view, it is interesting to see what kind of functions appear as solutions of the relevant canonical differential equations and study their structure. This will give us a glimpse of the minimum mathematical complexities and difficulties we should expect when we consider their two-loop counterparts. From a phenomenological standpoint, these one-loop integrals will be required for the computation of two-loop corrections for $2\to3$ scattering processes involving more than one massive external particle and/or internal massive particles. 

\begin{figure}[]
    \centering
    \begin{tabular}{cccc}
    \subfloat[C]{\includegraphics[width=3.5cm]{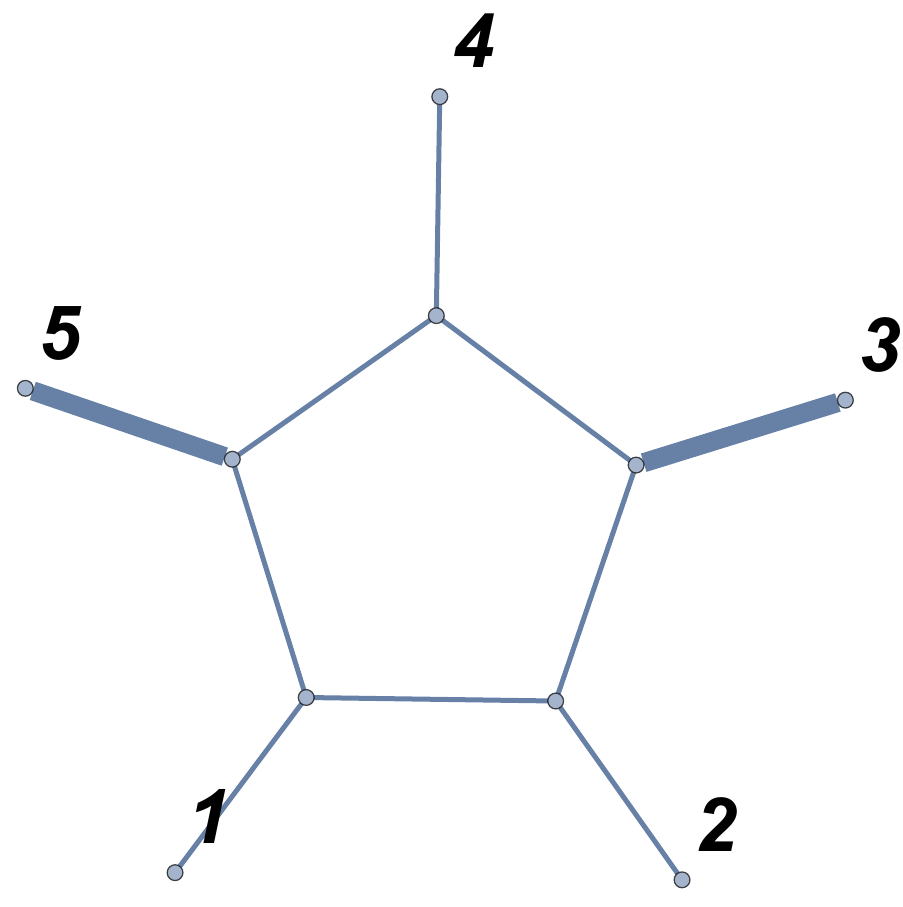}} &
    \subfloat[E]{\includegraphics[width=3.5cm]{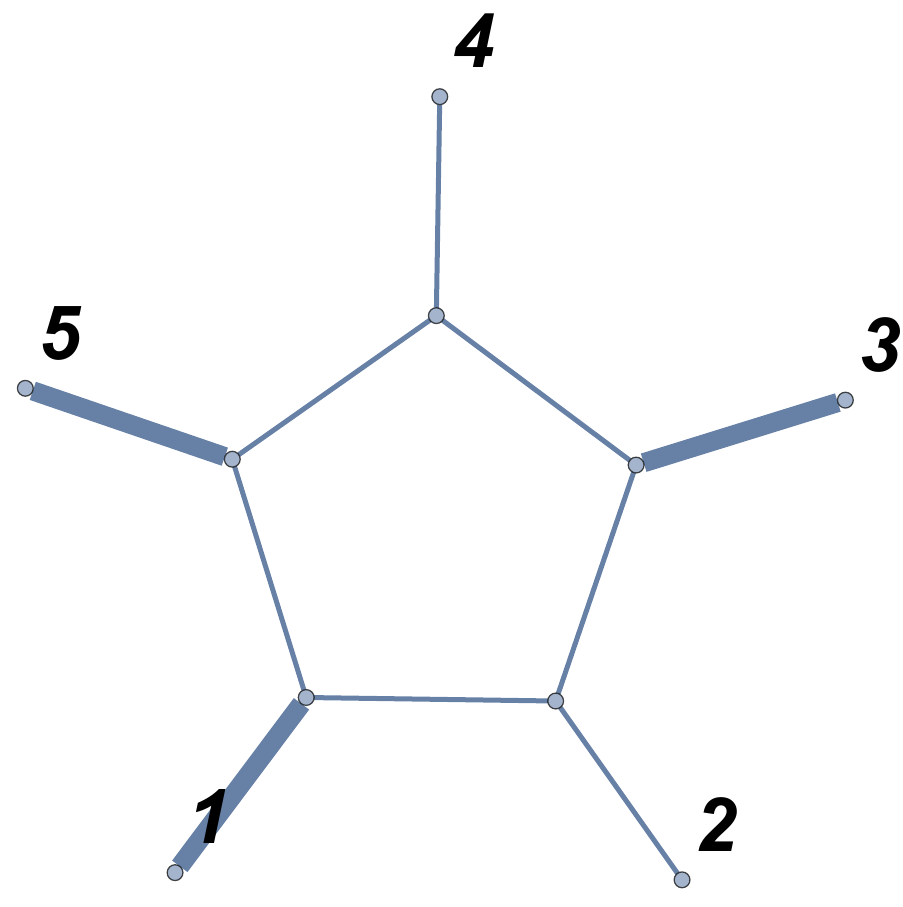}} &
    \subfloat[G]{\includegraphics[width=3.5cm]{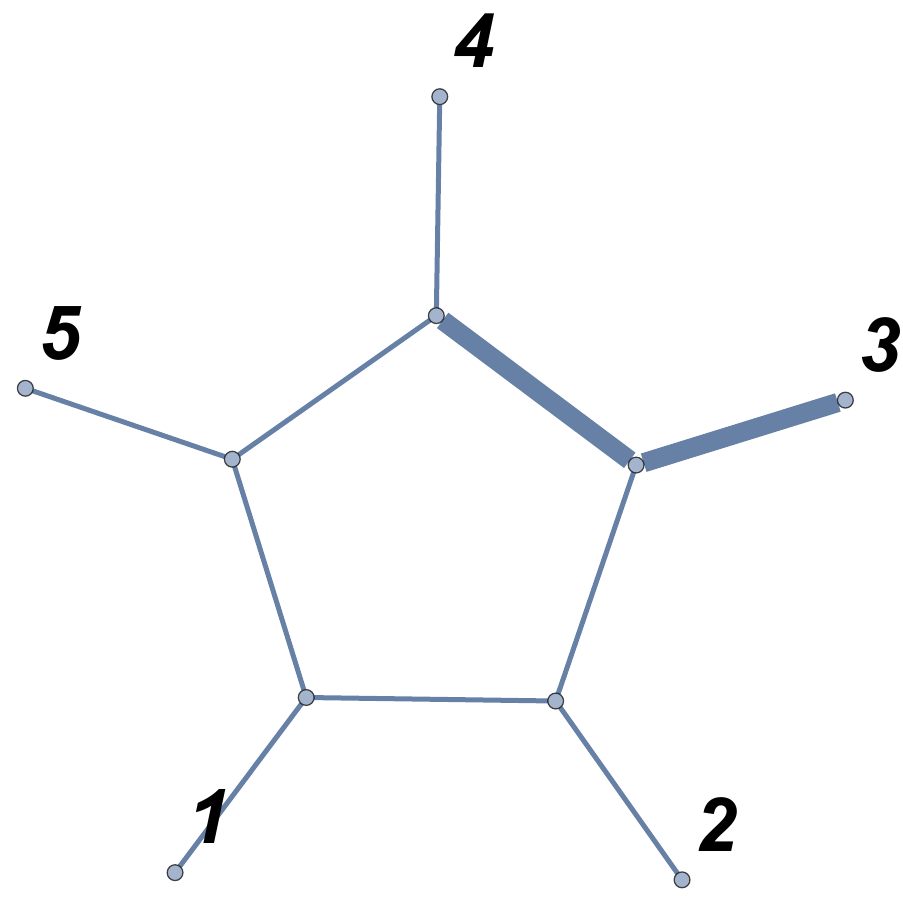}} &
    \subfloat[H]{\includegraphics[width=3.5cm]{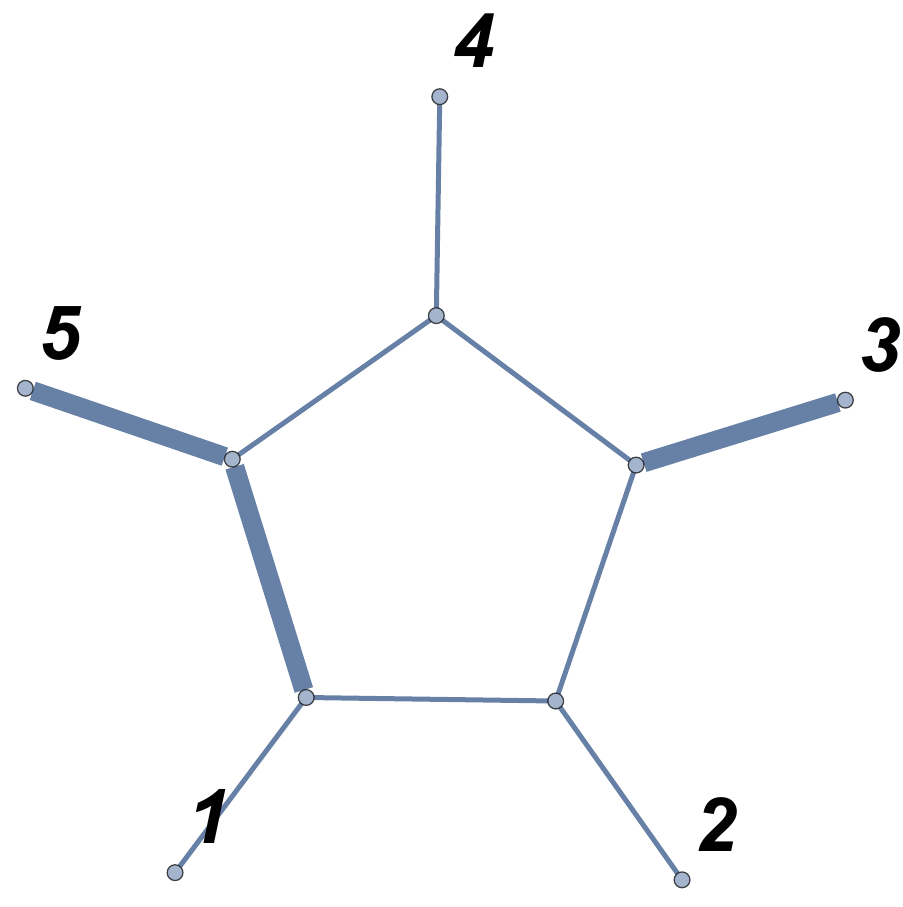}}
    \end{tabular}
    \caption{Top-sector diagrams for families computed with the SDE approach. All external particles are incoming. Bold external (internal) lines represent massive particles.}
    \label{fig:sdepenta}
\end{figure}

\begin{figure}[]
    \centering
    \begin{tabular}{cc}
    \subfloat[D]{\includegraphics[width=4cm]{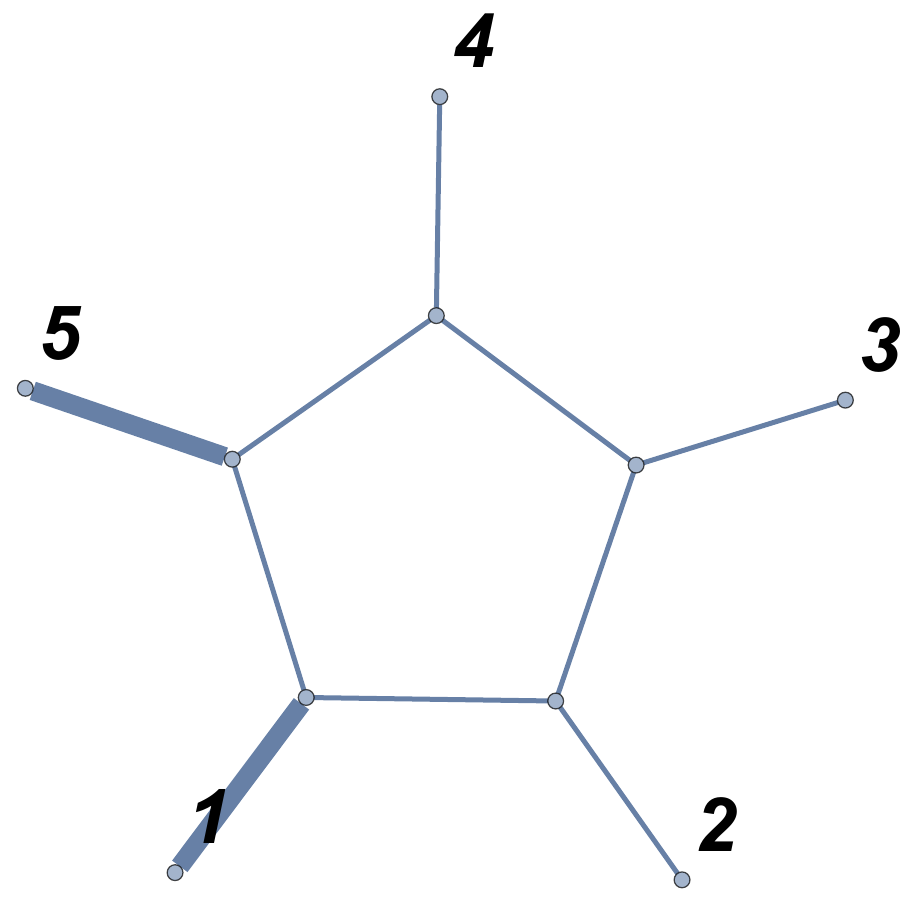}} &
    \subfloat[F]{\includegraphics[width=4cm]{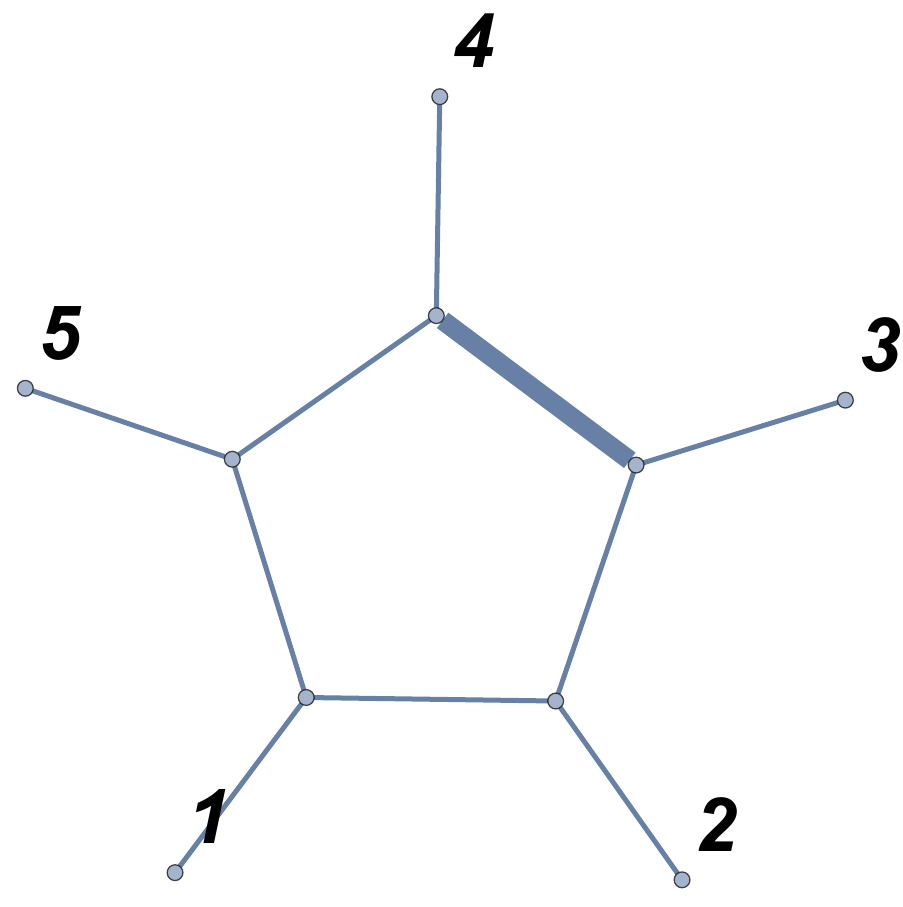}}
    \end{tabular}
    \caption{Top-sector diagrams for families computed through the $x\to1$ limit. All external particles are incoming. Bold external (internal) lines represent massive particles.}
    \label{fig:x1penta}
\end{figure}

In line with the arguments presented above, we consider in this paper the analytical calculation of several multiscale one-loop five-point Feynman integrals. More specifically, we present analytical results in terms of Goncharov polylogarithms of up to transcendental weight four for families with two and three off-shell legs and massless internal lines, as well as for families with one massive propagator and up to two external massive particles. Our calculation is based on the Simplified Differential Equations approach \cite{Papadopoulos:2014lla} which introduces an external dimensionless parameter $x$ in such a way that captures the off-shellness of one massive leg. The system of canonical differential equations is constructed by differentiating a pure basis of Master integrals in terms of $x$, regardless of the number of scales involved in the problem. One special feature of this approach is that by taking the limit of $x\to1$ \cite{Papadopoulos:2015jft} for a family with $n$ massive legs, we can obtain the result for a family with $n-1$ massive legs in an algorithmic way \cite{Canko:2020gqp}. In Figure \ref{fig:sdepenta} we present the families of Feynman integrals computed through the solution of simplified differential equations in canonical form, while in Figure \ref{fig:x1penta} we present the families of integrals computed through the  $x\to1$ limit.


The rest of our paper is structured as follows: in section \ref{sec:kin} we introduce basic notation and the kinematic configuration for each of the studied families of Feynman integrals, in section \ref{sec:results} we construct pure bases and derive and solve simplified differential equations in canonical form for all integral families depicted in Figure \ref{fig:sdepenta}, we present some of the resulting alphabets in $x$ and study their structure, and solve all integral families depicted in Figure \ref{fig:x1penta} through the $x\to1$ limit in terms of Goncharov polylogarithms of up to transcendental weight four. In section \ref{sec:valid} we provide an analysis of our results, as well as numerical checks for Euclidean points and in section \ref{sec:conclusions} we summarise our findings and discuss their key features. To the best of our knowledge these families have never before been considered in the literature, thus their solution constitutes an original contribution. This is also the first time that a calculation with the Simplified Differential Equations approach involving Master integrals with internal masses is reported. Along with this paper we provide all of our results in ancillary files. Explicit weight-three expressions for the top-sector basis elements of families $C$ and $H$ are given in appendix \ref{sec:w3result}.

\section{Notation and kinematics}
\label{sec:kin}

The integral families are defined through the following parametrization,
\begin{equation}\label{eq:pentaf}
    G_{a_1a_2a_3a_4a_5} = \int~ \frac{\mathrm{d}^d k_1}{i\pi^{(d/2)}}~ \frac{\mathrm{e}^{\epsilon \gamma_{E}}}{\mathcal{D}_1^{a_1}\mathcal{D}_2^{a_2}\mathcal{D}_3^{a_3}\mathcal{D}_4^{a_4}\mathcal{D}_5^{a_5}},\quad d=4-2\epsilon
\end{equation}
with
\begin{align}\label{eq:props}
    &\mathcal{D}_1=(k_1)^2-n_1~m^2,~ \mathcal{D}_2=(k_1+q_1)^2,~ \mathcal{D}_3=(k_1+q_1+q_2)^2 \nonumber \\
    &\mathcal{D}_4=(k_1+q_1+q_2+q_3)^2-n_4~m^2,~ \mathcal{D}_5=(k_1+q_1+q_2+q_3+q_4)^2
\end{align}
For the families $C, D$ and $E$ we have $n_1=n_4=0$, for the families $F$ and $G$ $n_1=0,n_4=1$ and finally for the family $H$ $n_1=1,n_4=0$. The kinematics for the families depicted in Figure \ref{fig:sdepenta} is as follows,
\begin{itemize}
    \item $C\&H$: $\sum_{i=1}^5q_i=0$, $q_i^2=0$, $i=1,2,4$, $q_3^2=m_3^2,q_5^2=m_5^2$
    \item $E$: $\sum_{i=1}^5q_i=0$, $q_i^2=0$, $i=2,4$, $q_1^2=\bar{m}_1^2,q_3^2=m_3^2,q_5^2=m_5^2$
    \item $G$: $\sum_{i=1}^5q_i=0$, $q_i^2=0$, $i=1,2,4,5$, $q_3^2=m_3^2$
\end{itemize}
We introduce the following $x$-parametrization\footnote{We use the abbreviations $p_{ij}=p_i+p_j$ and $p_{ijk}=p_i+p_j+p_k$ and similarly for $q$ later.}
\begin{equation}\label{eq:xparm}
    q_1=xp_1,~q_2=xp_2,~q_3=p_{123}-xp_{12},~q_{4}=p_{4},~q_5=-p_{1234}
\end{equation}
The kinematics in this underline momentum parametrization is
\begin{itemize}
    \item $C\&H$: $\sum_{i=1}^5p_i=0$, $p_i^2=0$, $i=1,2,3,4$, $p_5^2=m_5^2$
    \item $E$: $\sum_{i=1}^5p_i=0$, $p_i^2=0$, $i=2,3,4$, $p_1^2=m_1^2,p_5^2=m_5^2$
    \item $G$: $\sum_{i=1}^5p_i=0$, $p_i^2=0$, $i=1,2,3,4,5$
\end{itemize}
Introducing \eqref{eq:xparm} results in a mapping between the kinematic invariants in the original momentum parametrization, $q_i$, and the underline momentum parametrization $\{x,p_i\}$ for each of the families $C, E, G, H$.\footnote{We use the abbreviations $s_{ij}=q_{ij}^2,~ S_{ij}=p_{ij}^2$.}
\begin{align}\label{eq:mommap}
    C\&H:&~ m_3^2= (x-1) \left(S_{12} x-S_{45}\right),s_{12}= S_{12} x^2,s_{23}= S_{23}
   x-S_{45} x+S_{45}\nonumber \\
   &s_{34}= m_5^2 (-x)+m_5^2+x \left(S_{12}(x-1)+S_{34}\right),s_{45}=S_{45},s_{15}= m_5^2 (-x)+m_5^2+S_{15} x \nonumber \\
   E:&~ s_{12}= S_{12} x^2,s_{23}= x \left(m_1^2 (x-1)+S_{23}\right)-S_{45}
   x+S_{45},\nonumber \\
   &s_{34}= m_5^2 (-x)+m_5^2+x \left(S_{12} (x-1)+S_{34}\right),
   s_{45}=
   S_{45},\nonumber\\
   &s_{15}= x \left(m_1^2 (x-1)+S_{15}\right)+m_5^2 (-x)+m_5^2,\bar{m}_1^2= m_1^2
   x^2,m_3^2= (x-1) \left(S_{12} x-S_{45}\right) \nonumber \\
   G:&~ s_{12}= S_{12} x^2,s_{23}= S_{23} x-S_{45} x+S_{45},s_{34}= x \left(S_{12}
   (x-1)+S_{34}\right),s_{45}= S_{45},\nonumber \\
   &s_{15}= S_{15} x,m_3^2= (x-1) \left(S_{12}
   x-S_{45}\right)
\end{align}
For the families depicted in Figure \ref{fig:x1penta} their definition through \eqref{eq:pentaf}\&\eqref{eq:props} is obtained by taking \eqref{eq:xparm} and setting $x=1$, and their kinematic configuration is effectively the one produced by the underline momentum parametrization of the families through which we will calculate them with the $x\to1$ limit, therefore we have
\begin{itemize}
    \item $D~(x\to1~ \text{of}~ E)$: $\sum_{i=1}^5p_i=0$, $p_i^2=0$, $i=2,3,4$, $p_1^2=m_1^2,p_5^2=m_5^2$
    \item $F~(x\to1~ \text{of}~ G)$: $\sum_{i=1}^5p_i=0$, $p_i^2=0$, $i=1,2,3,4,5$
\end{itemize}

\section{Differential equations and pure solutions}
\label{sec:results}
In this section we will describe the analytical calculation of the integral families considered in this paper. We will construct pure bases for the families $C, E, G, H$ and use the Simplified Differential Equations approach \cite{Papadopoulos:2014lla} to compute them in terms of Goncharov polylogarithms. We will show how to obtain boundary terms \cite{Canko:2020ylt,Canko:2020gqp} for these canonical differential equations and present explicit results up to transcendental weight four. A discussion on the structure of the alphabets in $x$ for families $C$ and $H$ is also provided, along with their explicit expressions. 

We will also show how by taking the $x\to1$ limit of the analytic solution of families $E$ and $G$, one can obtain in an algorithmic way analytic results in terms of Goncharov polylogarithms for the families $D$ and $F$ respectively. Additionally, this procedure will allow us to obtain pure bases for the families $D$ and $F$ in a straightforward manner \cite{Canko:2020gqp}. The results presented here for these last two families are up to transcendental weight four as well.

It is important to note that introducing one internal mass does not appear to have any effect on the efficiency of the methods that have been developed for the determination of boundary terms, as well as taking the $x\to1$ limit within the Simplified Differential Equations approach.  

\subsection{Families $C, E, G, H$}
Constructing pure bases for the families $C, E, G, H$ is by now a trivial exercise. Following the reasoning of \cite{Abreu:2018rcw, Abreu:2020jxa}, the top sector basis element at the \textit{integrand} level is of the form
\begin{equation}\label{eq:uttop}
    \epsilon^2 \frac{\mathcal{P}_{11111}}{\sqrt{\Delta_5}} \Tilde{G}_{11111}
\end{equation}
where $\mathcal{P}_{11111}$ is the Baikov polynomial corresponding to the top sector \textit{integral} $G_{11111}$ for each family, $\Tilde{G}_{11111}$ is the top sector \textit{integrand} of each family  and $\Delta_5 = \det[q_i\cdot q_j]$ is the Gram determinant of the external momenta. The remaining pure basis elements can be constructed through the study of the leading singularities of their corresponding diagrams \cite{Henn:2014qga}. Using Azurite \cite{Georgoudis:2016wff} and Kira2 \cite{Klappert:2020nbg} we identify 15, 18, 16 and 18 Master Integrals for the families $C, E, G, H$ respectively. 

When considering five-point scattering, a number of square roots of the kinematic invariants enter the differential equations of the corresponding pure bases. These square roots originate from leading singularities of triangles with three massive legs\footnote{For all integral families considered in this paper the most complicated triangle Feynman integrals are the ones with fully massive legs and one massive propagator. This one internal mass however has no effect in the calculation of the leading singularity of the corresponding integral.} which are represented by square roots of the K\"allen function $\lambda(x,y,z) = x^2-2 x y-2 x z+y^2-2 y z+z^2$ and from square roots of the Gram determinants of the five-point external momenta. The existence of these square roots poses a challenge if one wishes to solve the differential equations analytically in terms of Goncharov polylogarithms. 

For the families considered in this subsection the following square roots appear:
\begin{align}
r_1&=\sqrt{\lambda(s_{12},m_3^2,s_{45})} \\
r_2&=\sqrt{\lambda(s_{12},s_{34},m_5^2)} \\
r_3&=\sqrt{\lambda(\bar{m}_1^2,s_{23},s_{45})} \\
r_4&=\sqrt{\lambda(\bar{m}_1^2,m_5^2,s_{15})} \\
r_5&=\sqrt{\Delta_5^C}=\sqrt{\Delta_5^H} \\
r_6&=\sqrt{\Delta_5^E} \\
r_7&=\sqrt{\Delta_5^G}
\end{align}
We should note at this point that not all square roots appear in every family at the same time.  More specifically, in families $C$ and $H$ we encounter $r_1, r_2, r_5$, in family $E$ $r_1, r_2, r_3, r_4, r_6$ and in family $G$ $r_1, r_7$. 

If one tries to compute these families using the standard differential equations approach, i.e. by differentiating with respect to all kinematic invariants, then the algebraic structure of the alphabet (i.e. the square roots appearing as letters) of the canonical differential equation for each family prohibits a straightforward solution in terms of Goncharov polylogarithms. In order to achieve a result in such a form we need to find a way to deal with these square roots. Several ideas have been put forward recently that are able to circumvent this problem and provide solutions in terms of Goncharov polylogarithms \cite{Heller:2019gkq,Besier:2018jen,Besier:2019kco,Bonetti:2020hqh} for specific cases, however a universal method to treat the problem of square roots appearing in the alphabet of canonical differential equations for multiscale families of Feynman integrals is still missing.

It turns out that for the families $C,E,G,H$ the $x$-parametrization introduced in \eqref{eq:xparm} and the resulting mapping of the kinematic invariants \eqref{eq:mommap} rationalises all square roots with respect to $x$. This allows us to obtain a canonical differential equation in $x$ for each of the four families considered in this subsection,
\begin{equation}\label{eq:cande}
\partial_{x} \textbf{g}=\epsilon \left( \sum_{i=1}^{l_{max}} \frac{\textbf{M}_i}{x-l_i} \right) \textbf{g}
\end{equation}
where $\textbf{g}$ is the pure basis for each family, $\textbf{M}_i$ are the residue matrices corresponding to each letter $l_i$ and $l_{max}$ is the length of the alphabet\footnote{Note here that we are following the notation of \cite{Canko:2020ylt,Syrrakos:2020kba,Canko:2020gqp} when talking about the \textit{letters} of the \textit{alphabet}.}. The kinematic dependence is entirely contained within the letters $l_i$, leaving the residue matrices $\textbf{M}_i$ to be solely constructed by rational numbers. The length of the alphabet for each of the four families considered in this subsection is $l_{max}^C=14,~l_{max}^E=19,~l_{max}^G=22,~l_{max}^H=30$. The explicit form of the alphabet for each of the four families is provided in the ancillary files that accompany this paper. In the next subsection we will study more closely some of these alphabets.

In order to solve \eqref{eq:cande} we need to provide boundary terms. We will follow closely the computational framework developed in \cite{Canko:2020ylt, Canko:2020gqp} for the determination of the relevant boundary terms. We start with the residue matrix corresponding to the letter $\{0\}$, $\textbf{M}_1$ and through its Jordan Decomposition we rewrite it as follows,
\begin{equation}
    \textbf{M}_1 =\textbf{S} \textbf{D} \textbf{S}^{-1}
\end{equation}
Then we define the \textit{resummation matrix} $\textbf{R}$ as follows
\begin{equation}\label{eq:resm0}
    \textbf{R} =\textbf{S} e^{\epsilon \textbf{D} \log(x)} \textbf{S}^{-1}
\end{equation}
The next step is to use IBP identities to write the pure basis $\textbf{g}$ in the following form 
\begin{equation}\label{eq:gtomasters}
    \textbf{g}=\textbf{T}\textbf{G}
\end{equation}
The list of Feynman integrals $\textbf{G}$ is also provided in electronic form. Furthermore, using the expansion-by-regions method \cite{Jantzen:2012mw} implemented in the \texttt{asy} code which is shipped along with \texttt{FIESTA4}~\cite{Smirnov:2015mct}, we can obtain information for the asymptotic behaviour of the Feynman integrals in terms of which we express the pure basis of Master integrals \eqref{eq:gtomasters} in the limit $x\to0$,
\begin{equation}\label{eq:regions}
    {G_i}\mathop  = \limits_{x \to 0} \sum\limits_j x^{b_j + a_j \epsilon }G^{(b_j + a_j \epsilon)}_{i} 
\end{equation}
where $a_j$ and $b_j$ are integers and $G_i$ are the individual members of the basis $\textbf{G}$ of Feynman integrals in \eqref{eq:gtomasters}. As explained in \cite{Canko:2020ylt}, we can construct the relation
\begin{equation}\label{eq:bounds}
    \mathbf{R} \mathbf{b}=\left.\lim _{x \rightarrow 0} \mathbf{T} \mathbf{G}\right|_{\mathcal{O}\left(x^{0+a_{j} \epsilon}\right)}
\end{equation}
where $\textbf{b}=\sum_{i=0}^n\epsilon^i\textbf{b}_0^{(i)}$ are the boundary terms that we need to compute. The right-hand-side of  \eqref{eq:bounds} implies that, apart from the terms $x^{a_i  \epsilon}$ coming from \eqref{eq:regions}, we expand around $x=0$, keeping only terms of order $x^0$.

Equation \eqref{eq:bounds} allows us to fix all the necessary boundary terms without the need of any further computation for the families $C,E,H$ while for family $G$ a few regions had to be computed. Similarly to \cite{Syrrakos:2020kba}, the resulting boundary terms for all of the four families considered in this subsection are in closed form, including some $_2F_1$ Hypergeometric functions which can be easily expanded to arbitrary powers of the dimensional regulator using \texttt{HypExp}~\cite{Huber:2007dx}. Therefore we are able to trivially obtain solutions of \eqref{eq:cande} for the families $C,E,G,H$ in terms of Goncharov polylogarithms of arbitrary weight.

In this paper we present explicit results for the families $C,E,G,H$ in terms of Goncharov polylogarithms of up to transcendental weight four, which can be written in the following compact form,
\begin{align}\label{eq:solution}
   \textbf{g}&= \epsilon^0 \textbf{b}^{(0)}_{0} + \epsilon \bigg(\sum\mathcal{G}_{a}\textbf{M}_{a}\textbf{b}^{(0)}_{0}+\textbf{b}^{(1)}_{0}\bigg) \nonumber \\
   &+ \epsilon^2 \bigg(\sum\mathcal{G}_{ab}\textbf{M}_{a}\textbf{M}_{b}\textbf{b}^{(0)}_{0}+\sum\mathcal{G}_{a}\textbf{M}_{a}\textbf{b}^{(1)}_{0}+\textbf{b}^{(2)}_{0}\bigg) \nonumber \\
   &+ \epsilon^3 \bigg(\sum\mathcal{G}_{abc}\textbf{M}_{a}\textbf{M}_{b}\textbf{M}_{c}\textbf{b}^{(0)}_{0}+\sum\mathcal{G}_{ab}\textbf{M}_{a}\textbf{M}_{b}\textbf{b}^{(1)}_{0}+\sum\mathcal{G}_{a}\textbf{M}_{a}\textbf{b}^{(2)}_{0}+\textbf{b}^{(3)}_{0}\bigg) \nonumber \\
   &+ \epsilon^4 \bigg(\sum\mathcal{G}_{abcd}\textbf{M}_{a}\textbf{M}_{b}\textbf{M}_{c}\textbf{M}_{d}\textbf{b}^{(0)}_{0}+\sum\mathcal{G}_{abc}\textbf{M}_{a}\textbf{M}_{b}\textbf{M}_{c}\textbf{b}^{(1)}_{0}\nonumber \\
   &+ \sum\mathcal{G}_{ab}\textbf{M}_{a}\textbf{M}_{b}\textbf{b}^{(2)}_{0}+\sum\mathcal{G}_{a}\textbf{M}_{a}\textbf{b}^{(3)}_{0}+\textbf{b}^{(4)}_{0}\bigg)
\end{align}
were $\mathcal{G}_{ab\ldots}:= \mathcal{G}(l_a,l_b,\ldots;x)$ represent the Goncharov polylogarithms. The $b_0^{(i)}$ terms, with $i$ indicating the corresponding weight, consist of Zeta functions $\zeta(i)$, logarithms and Goncharov polylogarithms of weight $i$ which have as arguments rational functions of the underline kinematic variables $S_{ij}$.

Our results are presented in such a way that each coefficient of $\epsilon^k$ has transcendental weight $k$. If we assign weight $-1$ to $\epsilon$, then \eqref{eq:solution} has uniform weight zero. The closed-form expressions of the boundary terms trivialise the extension of \eqref{eq:solution} to higher transcendental weights (or higher orders in $\epsilon$).

\subsection{\boldmath The alphabet in $x$}
\label{subsec:alphabets}
It is instructive to have a closer look at the alphabets for some of these families and see what lessons can be learned. We will study the alphabets of families $C$ and $H$. We choose these families because they have the same external kinematics but differ on the fact that family $H$ has one internal mass. Thus it is interesting to see how the introduction of an internal mass effects the alphabet. The conclusions drawn from the study of these families' alphabets are similar to what can be learned from the alphabets of the rest of the families considered in this paper.

\subsubsection*{Family $C$}
The alphabet for this family is
\begin{align}\label{eq:alphabetC}
    &l_1\to 0,l_2\to 1,l_3\to \frac{m^2_5}{m^2_5-S_{15}},l_4\to \frac{S_{12}+S_{23}}{S_{12}},l_5\to \frac{S_{12}+S_{15}-S_{34}}{S_{12}},\nonumber\\
    &l_6\to \frac{-\sqrt{\Delta
   _1}+m^2_5+S_{12}-S_{34}}{2 S_{12}},l_7\to \frac{\sqrt{\Delta _1}+m^2_5+S_{12}-S_{34}}{2 S_{12}},l_8\to \frac{m^2_5-S_{45}}{m^2_5-S_{15}+S_{23}-S_{45}},\nonumber\\
   &l_9\to
   \frac{m^2_5-S_{45}}{m^2_5-S_{34}-S_{45}},l_{10}\to \frac{S_{45}}{S_{12}},l_{11}\to \frac{S_{45}}{S_{45}-S_{23}},l_{12}\to \frac{m^2_5 S_{12}-S_{45}
   S_{12}+S_{34} S_{45}}{m^2_5 S_{12}-S_{12} S_{45}},\nonumber\\
   &l_{13}\to -\frac{\sqrt{\Delta _2}-2 m^2_5 S_{12}-m^2_5 S_{23}+S_{15} S_{12}-S_{23} S_{12}+2 S_{45}
   S_{12}+S_{23} S_{34}+S_{15} S_{45}-S_{34} S_{45}}{2 S_{12} \left(m^2_5-S_{15}+S_{23}-S_{45}\right)},\nonumber\\
   &l_{14}\to \frac{\sqrt{\Delta _2}+2 m^2_5 S_{12}+m^2_5
   S_{23}-S_{15} S_{12}+S_{23} S_{12}-2 S_{45} S_{12}-S_{23} S_{34}-S_{15} S_{45}+S_{34} S_{45}}{2 S_{12} \left(m^2_5-S_{15}+S_{23}-S_{45}\right)}
\end{align}
We notice that two square roots appear, $\sqrt{\Delta _1},\,\sqrt{\Delta _2}$, where $\Delta _1,\,\Delta _2$ are given by the following expressions
\begin{align}
    \Delta _1 =& -2 m^2_5 \left(S_{12}+S_{34}\right)+\left(m^2_5\right)^2+\left(S_{12}-S_{34}\right){}^2,\\
    \Delta _2 =& \left(m^2_5 S_{23}-S_{34}
   S_{23}+\left(S_{34}-S_{15}\right) S_{45}\right){}^2+S_{12}^2 \left(S_{15}-S_{23}\right){}^2\nonumber\\
   &+2 S_{12} \big(m^2_5 S_{23} \left(S_{15}-S_{23}-2 S_{34}\right)-S_{45} S_{15}^2+S_{23} S_{34}
   S_{15}+\left(S_{23}+S_{34}\right) S_{45} S_{15}\nonumber\\
   &+S_{23} S_{34} \left(S_{45}-S_{23}\right)\big)
\end{align}
These square roots are directly associated with $r_2$ and $r_5$ respectively. We would expect a third square root to appear, namely $r_1$, as mentioned in the previous subsection. However, after introducing the $x$-parametrization \eqref{eq:xparm}, $r_1$ is rational in all variables. More specifically, if we express these square roots using the first set of relations from \eqref{eq:mommap}, we get
\begin{align}
    r_1 =& \left(S_{12}-S_{45}\right) x\\
    r_2 =& \sqrt{\Delta _1}\, x\\
    r_5 =& \sqrt{\Delta _2}\, x^2
\end{align}
The structure of this alphabet is similar in terms of its complexity with the alphabet for the one-mass pentagon studied in \cite{Syrrakos:2020kba}.  The difference of course is the presence of an additional square root of the \textit{underline kinematic variables} $S_{ij}$ in the present case. 

\subsubsection*{Family $H$}
The alphabet for this family is
\begin{align}\label{eq:alphabetH}
    &l_1\to 0,l_2\to 1,l_3\to \frac{m^2}{S_{12}},l_4\to -\frac{\sqrt{m^2}}{\sqrt{S_{12}}},l_5\to \frac{\sqrt{m^2}}{\sqrt{S_{12}}},l_6\to
   \frac{S_{12}+S_{23}}{S_{12}},\nonumber\\
   &l_7\to \frac{m^2_5+S_{12}-S_{34}-\sqrt{\hat{\Delta} _1}}{2 S_{12}},l_8\to \frac{m^2_5+S_{12}-S_{34}+\sqrt{\hat{\Delta} _1}}{2 S_{12}},l_9\to
   \frac{m^2 \left(m^2_5+S_{12}-S_{34}-\sqrt{\hat{\Delta} _1}\right)}{2 m^2_5 S_{12}},\nonumber\\
   &l_{10}\to \frac{m^2 \left(m^2_5+S_{12}-S_{34}+\sqrt{\hat{\Delta} _1}\right)}{2 m^2_5
   S_{12}},l_{11}\to -\frac{m^2}{S_{23}-S_{45}},l_{12}\to \frac{m^2_5-S_{45}}{m^2_5-S_{34}-S_{45}},l_{13}\to \frac{m^2}{S_{45}},\nonumber\\
   &l_{14}\to
   \frac{S_{45}}{S_{12}},l_{15}\to -\frac{S_{45}}{S_{23}-S_{45}},l_{16}\to \frac{-S_{12} m^2-S_{12} S_{45}+\sqrt{\hat{\Delta} _2}}{2 S_{12}
   \left(S_{23}-S_{45}\right)},l_{17}\to -\frac{S_{12} m^2+S_{12} S_{45}+\sqrt{\hat{\Delta} _2}}{2 S_{12} \left(S_{23}-S_{45}\right)},\nonumber\\
   &l_{18}\to -\frac{-m^2_5 m^2+S_{34}
   m^2+S_{45} m^2-m^2_5 S_{12}+S_{12} S_{45}-S_{34} S_{45}+\sqrt{\hat{\Delta} _3}}{2 S_{12} \left(m^2_5-S_{45}\right)},\nonumber\\
   &l_{19}\to \frac{m^2_5 m^2-S_{34} m^2-S_{45}
   m^2+m^2_5 S_{12}-S_{12} S_{45}+S_{34} S_{45}+\sqrt{\hat{\Delta} _3}}{2 S_{12} \left(m^2_5-S_{45}\right)},\nonumber\\
   &l_{20}\to \frac{m^2}{m^2_5-S_{15}},l_{21}\to
   \frac{m^2_5}{m^2_5-S_{15}},l_{22}\to \frac{m^2_5-S_{45}}{m^2_5-S_{15}+S_{23}-S_{45}},l_{23}\to \frac{S_{12}+S_{15}-S_{34}}{S_{12}},\nonumber\\
   &l_{24}\to \frac{m^2
   \left(m^2_5-S_{45}\right)}{\left(m^2_5-S_{15}+S_{23}-S_{45}\right) m^2-m^2_5 S_{23}+S_{15} S_{45}},l_{25}\to \frac{S_{12} m^2+m^2_5 S_{12}-\sqrt{\hat{\Delta} _4}}{2
   S_{12} \left(m^2_5-S_{15}\right)},\nonumber\\
   &l_{26}\to \frac{S_{12} m^2+m^2_5 S_{12}+\sqrt{\hat{\Delta} _4}}{2 S_{12} \left(m^2_5-S_{15}\right)},\nonumber\\
   &l_{27}\to \frac{-S_{12} S_{15}
   m^2+m^2_5 S_{23} m^2+S_{12} S_{23} m^2-S_{23} S_{34} m^2-S_{15} S_{45} m^2+S_{34} S_{45} m^2+\sqrt{\hat{\Delta} _5}}{2 S_{12} \left(m^2_5 S_{23}-S_{15}
   S_{45}\right)},\nonumber\\
   &l_{28}\to -\frac{S_{12} S_{15} m^2-m^2_5 S_{23} m^2-S_{12} S_{23} m^2+S_{23} S_{34} m^2+S_{15} S_{45} m^2-S_{34} S_{45} m^2+\sqrt{\hat{\Delta} _5}}{2
   S_{12} \left(m^2_5 S_{23}-S_{15} S_{45}\right)},\nonumber\\
   &l_{29}\to \frac{2 m^2_5 S_{12}-S_{15} S_{12}+S_{23} S_{12}-2 S_{45} S_{12}+m^2_5 S_{23}-S_{23} S_{34}-S_{15}
   S_{45}+S_{34} S_{45}+\sqrt{\hat{\Delta} _6}}{2 S_{12} \left(m^2_5-S_{15}+S_{23}-S_{45}\right)},\nonumber\\
   &l_{30}\to -\frac{-2 m^2_5 S_{12}+S_{15} S_{12}-S_{23} S_{12}+2 S_{45}
   S_{12}-m^2_5 S_{23}+S_{23} S_{34}+S_{15} S_{45}-S_{34} S_{45}+\sqrt{\hat{\Delta} _6}}{2 S_{12} \left(m^2_5-S_{15}+S_{23}-S_{45}\right)}
\end{align}
The first remark that we can make for this alphabet is that we have six square roots in the \textit{underline kinematic variables} $S_{ij}$, namely $\sqrt{\hat{\Delta} _i},\, i=1,\ldots 6$. The explicit expressions for the arguments of these square roots are as follows,
\begin{align}
    \hat{\Delta }_1=& -2 m_5^2 \left(S_{12}+S_{34}\right)+m_5^4+\left(S_{12}-S_{34}\right){}^2, \\\hat{\Delta }_2=& S_{12} \left(m^4 S_{12}+4 m^2 S_{23}
   \left(S_{12}+S_{23}\right)-2 m^2 \left(S_{12}+2 S_{23}\right) S_{45}+S_{12} S_{45}^2\right), 
   \\\hat{\Delta }_3=& m^4 \left(-m_5^2+S_{34}+S_{45}\right){}^2+\left(m_5^2 S_{12}+\left(S_{34}-S_{12}\right) S_{45}\right){}^2\nonumber \\
   &+2 m^2
   \big(m_5^2 \left(-S_{12}\right) \left(m_5^2+S_{34}\right)+m_5^2 \left(2 S_{12}+S_{34}\right) S_{45}-\left(S_{12}+S_{34}\right) S_{45}^2\nonumber \\
   &+\left(S_{12}-S_{34}\big)
   S_{34} S_{45}\right), 
   \\\hat{\Delta }_4=& S_{12} \left(m^4 S_{12}-2 m_5^2 m^2 \left(S_{12}+2 S_{15}-2
   S_{34}\right)+4 m^2 S_{15} \left(S_{12}+S_{15}-S_{34}\right)+m_5^4 S_{12}\right), \\\hat{\Delta }_5=& m^4\, \hat{\Delta }_6, 
   \\\hat{\Delta }_6=& \left(m_5^2
   S_{23}-S_{23} S_{34}+\left(S_{34}-S_{15}\right) S_{45}\right){}^2+S_{12}^2 \left(S_{15}-S_{23}\right){}^2\nonumber \\
   &+2 S_{12} \big(m_5^2 S_{23} \left(S_{15}-S_{23}-2 S_{34}\right)+S_{15} S_{23} S_{34}-S_{15}^2
   S_{45}+S_{15} \left(S_{23}+S_{34}\right) S_{45}\nonumber \\
   &+S_{23} S_{34} \left(S_{45}-S_{23}\right)\big)
\end{align}
In comparison with \eqref{eq:alphabetC}, we see the same two square roots associated with the leading singularities of the massive three-point functions, $r_2$ and with the Gram determinant of the external momenta, $r_5$, namely
\begin{align}
    r_2 =& \sqrt{\Delta _1}\, x = \sqrt{\hat{\Delta} _1}\, x\\
    r_5 =& \sqrt{\Delta _2}\, x^2 = \sqrt{\hat{\Delta} _6}\, x^2
\end{align}
We have however four more square roots, $\sqrt{\hat{\Delta} _2},\, \sqrt{\hat{\Delta} _3},\, \sqrt{\hat{\Delta} _4},\, \sqrt{\hat{\Delta} _5}$, which involve the internal mass $m^2$, and are not directly associated with any leading singularities of any diagram from this family. We see therefore that the introduction of an internal mass has a major impact on the complexity of the resulting alphabet.


\subsubsection*{Final remarks and comparison with other methods}
We have seen that introducing an internal mass can significantly increase the complexity of the algebraic structure of an alphabet. It would be interesting to find an explanation for the appearance of the additional square roots in \eqref{eq:alphabetH} which are not directly associated with a leading singularity. Normally, one would expect that the square roots that appear in the alphabet also appear in the definition of the pure basis elements, although in the case of family $H$ the extra square roots do not appear in the pure basis definition.

On a more general note, whenever the Simplified Differential Equations approach has been applied in conjunction with a pure basis \cite{Canko:2020ylt, Syrrakos:2020kba, Canko:2020gqp}, we have observed a \textit{reduction} in the number of letters that appear when compared with the alphabets that arise through the usual method of differential equations, i.e. when one differentiates with respect to all kinematic invariants. It would be interesting to see the structure of the alphabets for the families studied here when one uses the usual method of differential equations and whether this feature of alphabets in $x$ with fewer letters still holds.

\subsection{Families $D, F$}
For these families we will obtain analytic expressions through the results of the families $E$ and $G$. We will follow the procedure of taking the $x\to1$ limit of our solution for a family with $n$ massive legs to obtain a pure basis and analytic solution of a family with $n-1$ massive legs, as described in detail in \cite{Canko:2020gqp}.

For families $E$ and $G$ we exploit the shuffle properties of Goncharov polylogarithms to write their solution \eqref{eq:solution} as an expansion in terms of $\log(1-x)$ as follows
\begin{equation}\label{eq:solexp1}
    \textbf{g}=\sum_{n\geq 0} \epsilon^{n} \sum_{i=0}^{n}\frac{1}{i!}\textbf{c}^{(n)}_{i} \log^i(1-x)
\end{equation}
with all $\textbf{c}^{(n)}_{i}$ being finite in the limit $x\to 1$. The next step is to define the regular part of \eqref{eq:solexp1} at $x=1$
\begin{equation}\label{eq:reg1}
  \textbf{g}_{reg}=\sum_{n\geq 0}\epsilon^n \textbf{c}^{(n)}_{0}
\end{equation}
and after setting $x=1$ explicitly in \eqref{eq:reg1} we may define the truncated part of \eqref{eq:solexp1},
\begin{equation}\label{eq:trunc}
    \textbf{g}_{trunc} = \textbf{g}_{reg}(x=1)
\end{equation}
Having done that, we utilise the residue matrix that corresponds to the letter $\{1\}$, $\textbf{M}_2$, and define the \textit{resummation matrix} $\Tilde{\textbf{R}}$ as follows
\begin{equation}\label{eq:resum1}
    \Tilde{\textbf{R}}  = \Tilde{\textbf{S}} e^{\epsilon \Tilde{\textbf{D}} \log(1-x)} \Tilde{\textbf{S}}^{-1}
\end{equation}
were $\Tilde{\textbf{S}}, \Tilde{\textbf{D}}$ are constructed through the Jordan decomposition of $\textbf{M}_2$, i.e. $\textbf{M}_2=\Tilde{\textbf{S}} \Tilde{\textbf{D}} \Tilde{\textbf{S}}^{-1}$.
The \textit{resummation matrix} $\Tilde{\textbf{R}}$ has terms of $(1-x)^{a_i\epsilon}$, with $a_i$ being the eigenvalues of $\textbf{M}_2$. After setting all terms $(1-x)^{a_i\epsilon}$ equal to zero, we define the purely numerical matrix $\Tilde{\textbf{R}}_0$. Obtaining the $x\to1$ limit of \eqref{eq:solution} amounts to acting with $\Tilde{\textbf{R}}_0$ on \eqref{eq:trunc}
\begin{equation}\label{eq:glim1}
    \textbf{g}_{x\to 1} = \Tilde{\textbf{R}}_0 \textbf{g}_{trunc}
\end{equation}

Up to now we have calculated the $x\to1$ limit for families $E$ and $G$. This operation not only yields the result for a given family of Master integrals with $n$ massive legs at a special limit, but also allows us to obtain results for an independent family of Master integrals with $n-1$ massive legs. In the case of family $E$ for example, taking the $x\to1$ limit makes $q_3$ to become massless through \eqref{eq:xparm}, thus yielding the kinematics for family $D$. However, having the $x\to1$ limit of $E$ means that we have explicit solutions for the 18 pure basis elements of that family, whereas family $D$ has 16 basis elements. This means that out of the 18 basis elements of family $E$ at this limit, we need to find the 16 of them that form the pure basis and result for family $D$, i.e. at $x\to1$ only 16 of the 18 basis elements of family $E$ should remain linearly independent. The same reasoning holds for family $G$ which has 16 basis elements while family $F$ has 15.

In order to find the pure bases for families $D$ and $F$ we can either use Integration-By-Part identities to do the reduction or follow the approach described in \cite{Canko:2020gqp}. We shall use the latter method in the following. 

In all cases that we have considered so far, $\Tilde{\textbf{R}}_0$ is always an \textit{idempotent} matrix which means that, among others,  it has the following very useful property
\begin{equation}\label{eq:r10}
    \Tilde{\textbf{R}}_0^2=\Tilde{\textbf{R}}_0
\end{equation}
Acting with $\Tilde{\textbf{R}}_0$ on \eqref{eq:glim1} and using \eqref{eq:r10} yields the following relation
\begin{align}\label{eq:methodx1}
    \Tilde{\textbf{R}}_0 \textbf{g}_{x\to 1} &= \Tilde{\textbf{R}}_0^2 \textbf{g}_{trunc} \nonumber \\
    &=\Tilde{\textbf{R}}_0 \textbf{g}_{trunc} \nonumber \\
    &=\textbf{g}_{x\to 1}
\end{align}
This relation, solved as an equation for each row, produces relations that allow us to determine the linearly independent basis elements for families $D$ and $F$. Therefore, applying \eqref{eq:methodx1} to the pure basis of $E$ and $G$ yields in an algorithmic way the pure bases for families $D$ and $F$.

\subsection{On the choice of integral families}

The basic rule for choosing which five-point family to consider is to have the one-mass result \cite{Syrrakos:2020kba} as a starting point and add masses, with the condition that their simplified differential equations in canonical form have alphabets rational in $x$, when one uses parametrization \eqref{eq:xparm}. If the resulting alphabet is not rational in $x$, then calculating the integral family through the $x\to1$ limit of another is considered. The exception to the above rule is the five-point family with one internal mass and massless external legs, which in the framework of the Simplified Differential Equations can only be calculated through the $x\to1$ limit of its corresponding one-mass family. 

More specifically, if one tries to calculate family $D$ using \eqref{eq:xparm} and deriving differential equations in $x$, then the resulting alphabet is not rational in $x$. Nevertheless, family $D$ can be expressed in terms of Goncharov polylogarithms through the $x\to1$ limit of family $E$. Apart from family $E$ there is another family with massless propagators and three massive legs, the one which all three masses are adjacent. However, the alphabet of this family is not rational in $x$, if one uses \eqref{eq:xparm} to parametrize it. 

Introducing an internal mass allows for many more families to be considered. Apart from the ones presented in this paper, a family with one internal mass and three massive legs (i.e. take family $H$ and regard $q_1$ as massive) was considered, however its alphabet in $x$ is not rational using \eqref{eq:xparm}.

If the families with non-rational alphabets in $x$ can be rationalised using a parametrization other than \eqref{eq:xparm} remains an open question. 

\section{Validation}
\label{sec:valid}

For all families computed in this paper we have made heavy use of the \texttt{Mathematica} package \texttt{PolyLogTools} \cite{Duhr:2019tlz} for the manipulation of the resulting Goncharov polylogarithms. As shown in \eqref{eq:solution}, we provide explicit results up to order $\mathcal{O}(\epsilon^4)$. In Table \ref{tab:numgpl} we provide an analysis of our results for each family, regarding the number of Goncharov polylogarithms that appear in each transcendental weight, where the weight is counted as the number of $l_i$ indices of $\mathcal{G}(l_a,l_b,\ldots;x)$. These numbers are obtained by gathering all Goncharov polylogarithms that appear up to order $\mathcal{O}(\epsilon^4)$ in each integral family, and distinguishing them according to their corresponding weight. For comparison, we perform the same task for the top-sector basis element of each family. 

A common feature of our results is that our solutions are dominated by the number of weight-four polylogarithmic functions. Due to the universally transcendental feature of our solutions, the $\mathcal{O}(\epsilon^4)$ part is expected to be the most cumbersome to calculate numerically, since weight-four Goncharov polylogarithms take longer to calculate than lower-weight ones. To avoid any misconceptions, it should be noted that each top-sector basis element starts from $\mathcal{O}(\epsilon^3)$, despite containing lower-weight polylogarithms.

 In appendix \ref{sec:w3result} we present explicit formulas for the weight-three part of the pure top-sector basis elements of families $C$ and $H$ in order to give an idea of the structure and length of the resulting expressions, as well as the way the letters of the alphabets studied in subsection \ref{subsec:alphabets} are introduced in the relevant solutions.

\begin{table}[ht]
    \centering
\begin{tabular}{|c|c|c|c|c|c|}
\hline
 Family     & W=1 & W=2 & W=3 & W=4 & Total
 \\
 \hline 
$C$      & 9 (0)   & 54 (16) & 204 (106) & 605 (272) & 872 (394)
\\
\hline
$E$     & 13 (0)  & 87 (24) & 349 (172) & 1033 (432) & 1482 (628)
\\
\hline
$G$     & 21 (4)  & 163 (50) & 878 (329) & 2160 (884) & 3222 (1267)
\\
\hline
$H$     & 19 (0)  & 195 (42) & 1527 (616) & 5914 (2732) & 7655 (3390)
\\
\hline
$D$     & 11 (0)  & 83 (24) & 393 (192) & 1445 (656) & 1932 (872)
\\
\hline
$F$     & 19 (4)  & 151 (50) & 872 (349) & 2356 (1042) & 3398 (1445)
\\
\hline
\end{tabular}
\caption{Number of Goncharov polylogarithms entering the solution. Results for the respective top-sector basis elements are in parenthesis.}
\label{tab:numgpl}
\end{table}

Regarding the validation of our results, we have performed numerical checks of our solution for each family against \texttt{pySecDec} \cite{Borowka:2017idc} for Euclidean points. All Goncharov polylogarithms have been computed numerically using the \texttt{Ginsh} command of \texttt{PolyLogTools} \cite{Duhr:2019tlz}, as well as \texttt{handyG} \cite{Naterop:2019xaf}, which is a Fortran implementation of the algorithms developed in \cite{Vollinga:2004sn}. For all checks that we have performed we have found perfect agreement. In Table \ref{tab:numres} we also provide numerical results and timing for the top-sector basis element for each corresponding family. We include timings using \texttt{handyG} since we found that it is in general faster, although it is restrictive in its precision compared to \texttt{PolyLogTools}.

\begin{table}[ht]
    \centering
\begin{tabular}{|c|c|c|}
\hline
 Top-Sector     & Time (sec) & Result  \\
 \hline 
$C$      & 0.146897   & $-0.314547 \epsilon ^4-0.120811 \epsilon ^3$\\
\hline
$E$      & 0.248436   & $-0.0332408 \epsilon ^4-0.0215131 \epsilon ^3$ \\
\hline
$G$     & 0.475048  & $-0.439003 \epsilon ^4-0.130267 \epsilon ^3$ \\
\hline
$H$      & 1.89365   & $-0.0165223 \epsilon ^4-0.0192393 \epsilon ^3$\\
\hline
$D$      & 2.15734   & $-0.127286 \epsilon ^4-0.162439 \epsilon ^3$\\
\hline
$F$      & 0.730996   & $-0.528266 \epsilon ^4-0.33331 \epsilon ^3$\\
\hline
\end{tabular}
\caption{Numerical computation of Goncharov polylogarithms using \texttt{handyG} with double precision. The computations were performed on a 1,6 GHz Intel Core i5 laptop using a single CPU core.}
\label{tab:numres}
\end{table}

Along with this paper we provide all of our results in the ancillary files. More specifically, for the families $C,E, G$ and $H$, we provide the pure basis, boundary terms in closed form, the alphabet, the Feynman integrals in terms of which we have expressed the pure basis, the residue matrices for the canonical differential equation, as well as the explicit result for each of the aforementioned families in terms of Goncharov polylogarithms up to transcendental weight four. For families $D$ and $F$ we provide the pure basis, the alphabet and explicit results in terms of Goncharov polylogarithms up to transcendental weight four.
All ancillary files are in the form \textit{familyname.m}. We also provide for all families the Euclidean points that were used to obtain the numerical results of Table \ref{tab:numres} in the file \textit{EPoints.m}.

\section{Conclusions}
\label{sec:conclusions}

The current frontier in the calculation of multiscale multiloop Feynman integrals for $2\to3$ scattering processes relevant to LHC searches lies at two-loop five-point Feynman integrals with one off-shell leg and massless internal lines. As of this writing, results for all planar two-loop five-point Master integrals have been obtained using a numerical \cite{Abreu:2020jxa}, as well as an analytical approach \cite{Canko:2020ylt}. Regarding the analytical results, all planar Master integrals were expressed in terms of Goncharov polylogarithms of up to transcendental weight four. These results, along with analytic results for the relevant one-loop five-point Master integrals with one off-shell leg \cite{Syrrakos:2020kba}, were recently used to perform the first fully analytic calculation of a two-loop scattering amplitude for $Wb\bar{b}$ production \cite{Badger:2021nhg}. Furthermore, using a new method for calculating Master integrals, the authors of \cite{Papadopoulos:2019iam} have computed numerically one of the non-planar two-loop five-point families. More recently, some of the authors of \cite{Abreu:2020jxa} presented the three \textit{hexabox} topologies in \cite{Abreu:2021smk} using the same techniques as in \cite{Abreu:2020jxa}.

Looking ahead, at some point we will have to consider more complicated Feynman integrals, involving more massive external particles and/or massive propagators. One of the expected challenges when considering such integrals is the introduction of many square roots in the alphabet of the resulting canonical differential equations. It remains a non-trivial exercise to find a universal way to handle these roots and achieve a result in terms of Goncharov polylogarithms, however several ideas have been put forward in recent times \cite{Heller:2019gkq,Besier:2018jen,Besier:2019kco,Bonetti:2020hqh}. One should also keep in mind that even if a so-called \textit{dlog} form of the differential equations is achieved, it does not guarantee that its solution will be in terms of Goncharov polylogarithms \cite{Brown:2020rda}.

In order to get a glimpse of the complexities that lie beyond the frontier of five-point scattering involving one off-shell leg and massless internal lines, in this paper we studied families of one-loop five-point Feynman integrals with two and three massive external legs and massless propagators, as well as one-loop five-point families with one massive internal line and up to two massive external legs. 

We used the Simplified Differential Equations approach for the construction of canonical differential equations for pure bases of the families of Figure \ref{fig:sdepenta} and as a special limit we obtained results for the families of Figure \ref{fig:x1penta}. As it turned out, the parametrization \eqref{eq:xparm} was enough to rationalise all square roots introduced in the alphabet of families $C,E,G,H$ of Figure \ref{fig:sdepenta}. For these families we were also able to obtain boundary terms for the canonical differential equations in closed form, allowing us to trivially derive solutions for these families in terms of Goncharov polylogarithms of arbitrary transcendental weight.

For families $D,F$ of Figure \ref{fig:x1penta} we obtained analytic results through a special limit of our solutions for families $E$ and $G$ respectively. For all families studied in this paper we provide explicit results in terms of Goncharov polylogarithms of up to transcendental weight four.

Regarding the structure of the resulting alphabets in $x$, we saw that when one internal mass is introduced, square roots involving this mass arise, which are not present in the definition of the pure basis. Further study of these alphabets is required to pin-point the origin of these additional square roots, which we leave for future work. It is also interesting to explore in the future the structure of these alphabets when one employs the standard method of differential equations, i.e. differentiating with respect to all kinematic variables. A comparison between the two approaches might further elucidate the effectiveness of the Simplified Differential Equations approach in providing solutions to multiscale Feynman integrals in terms of Goncharov polylogarithms, as well as provide an idea of whether the representation of these integrals in terms of polylogarithmic functions is the best one for phenomenological applications.

\acknowledgments
The author acknowledges fruitful discussions with Costas G. Papadopoulos, Dhimiter D. Canko and Lorenzo Tancredi.

\appendix

\section{Explicit results at weight three}
\label{sec:w3result}
In this appendix we provide explicit results for families $C$ and $H$ for the weight-three part of each top-sector pure basis element. Assuming that each top-sector basis element is expressed in the following manner,
\begin{equation}
    g_{i}^\text{family} = \sum_{w=3}^4 \epsilon^w\, \Tilde{g}_{i,w}^\text{family}
\end{equation}
we give the explicit expression for the $\Tilde{g}_{i,3}^\text{family}$ part. We introduce the following shorthand notations for brevity,
\begin{align}
    &\mathcal{G}_{a,b,\ldots} = \mathcal{G}(l_a,l_b,\ldots;x)\\
    &L_1= \log \left(-S_{12}\right),L_2= \log \left(-S_{45}\right),L_3= \log \left(-m_5^2\right),\\
    &L_4= \log \left(m^2\right),L_5= \log
   \left(\frac{m^2-S_{45}}{m^2}\right),L_6= \log \left(\frac{m^2-m_5^2}{m^2}\right)
\end{align}
\subsection{Top sector of family C}
In the following formula note that letters $l_i,\, i=\{6,7,13,14\}$ contain square roots in the \textit{underline kinematic variables} $S_{ij}$, as shown explicitly in \eqref{eq:alphabetC}.

\begin{align}
   \Tilde{g}_{15,\,3}^C=&\frac{1}{4} L_1 \mathcal{G}_{13,2}-\frac{1}{4} L_2 \mathcal{G}_{13,2}+\frac{1}{4} L_1 \mathcal{G}_{13,3}-\frac{1}{4} L_3
   \mathcal{G}_{13,3}-\frac{1}{4} L_1 \mathcal{G}_{13,6}+\frac{1}{4} L_3 \mathcal{G}_{13,6}-\frac{1}{4} L_1 \mathcal{G}_{13,7}\nonumber \\
   &+\frac{1}{4} L_3
   \mathcal{G}_{13,7}-\frac{1}{4} L_2 \mathcal{G}_{13,8}+\frac{1}{4} L_3 \mathcal{G}_{13,8}+\frac{1}{4} L_2 \mathcal{G}_{13,9}-\frac{1}{4} L_3
   \mathcal{G}_{13,9}+\frac{1}{4} L_1 \mathcal{G}_{13,10}-\frac{1}{4} L_2 \mathcal{G}_{13,10}\nonumber \\
   &-\frac{1}{4} L_1
   \mathcal{G}_{13,11}+\frac{1}{4} L_2 \mathcal{G}_{13,11}-\frac{1}{4} L_1 \mathcal{G}_{14,2}+\frac{1}{4} L_2 \mathcal{G}_{14,2}-\frac{1}{4}
   L_1 \mathcal{G}_{14,3}+\frac{1}{4} L_3 \mathcal{G}_{14,3}+\frac{1}{4} L_1 \mathcal{G}_{14,6}\nonumber \\
   &-\frac{1}{4} L_3 \mathcal{G}_{14,6}+\frac{1}{4}
   L_1 \mathcal{G}_{14,7}-\frac{1}{4} L_3 \mathcal{G}_{14,7}+\frac{1}{4} L_2 \mathcal{G}_{14,8}-\frac{1}{4} L_3 \mathcal{G}_{14,8}-\frac{1}{4}
   L_2 \mathcal{G}_{14,9}+\frac{1}{4} L_3 \mathcal{G}_{14,9}\nonumber \\
   &-\frac{1}{4} L_1 \mathcal{G}_{14,10}+\frac{1}{4} L_2
   \mathcal{G}_{14,10}+\frac{1}{4} L_1 \mathcal{G}_{14,11}-\frac{1}{4} L_2 \mathcal{G}_{14,11}+\frac{1}{2}
   \mathcal{G}_{13,2,1}+\frac{1}{2} \mathcal{G}_{13,3,1}+\frac{1}{4} \mathcal{G}_{13,3,3}\nonumber \\
   &-\frac{1}{4} \mathcal{G}_{13,3,6}-\frac{1}{4}
   \mathcal{G}_{13,3,7}-\frac{1}{4} \mathcal{G}_{13,4,2}-\frac{1}{4} \mathcal{G}_{13,4,10}+\frac{1}{4}
   \mathcal{G}_{13,4,11}-\frac{1}{4} \mathcal{G}_{13,5,3}+\frac{1}{4} \mathcal{G}_{13,5,6}\nonumber \\
   &+\frac{1}{4} \mathcal{G}_{13,5,7}-\frac{1}{2}
   \mathcal{G}_{13,6,1}-\frac{1}{2} \mathcal{G}_{13,7,1}+\frac{1}{4} \mathcal{G}_{13,8,3}-\frac{1}{4} \mathcal{G}_{13,8,11}+\frac{1}{4}
   \mathcal{G}_{13,9,2}-\frac{1}{4} \mathcal{G}_{13,9,6}\nonumber \\
   &-\frac{1}{4} \mathcal{G}_{13,9,7}+\frac{1}{4} \mathcal{G}_{13,9,10}+\frac{1}{2}
   \mathcal{G}_{13,10,1}-\frac{1}{2} \mathcal{G}_{13,11,1}+\frac{1}{4} \mathcal{G}_{13,11,2}+\frac{1}{4}
   \mathcal{G}_{13,11,10}-\frac{1}{4} \mathcal{G}_{13,11,11}\nonumber \\
   &-\frac{1}{4} \mathcal{G}_{13,12,2}+\frac{1}{4}
   \mathcal{G}_{13,12,6}+\frac{1}{4} \mathcal{G}_{13,12,7}-\frac{1}{4} \mathcal{G}_{13,12,10}-\frac{1}{2}
   \mathcal{G}_{14,2,1}-\frac{1}{2} \mathcal{G}_{14,3,1}-\frac{1}{4} \mathcal{G}_{14,3,3}\nonumber \\
   &+\frac{1}{4} \mathcal{G}_{14,3,6}+\frac{1}{4}
   \mathcal{G}_{14,3,7}+\frac{1}{4} \mathcal{G}_{14,4,2}+\frac{1}{4} \mathcal{G}_{14,4,10}-\frac{1}{4}
   \mathcal{G}_{14,4,11}+\frac{1}{4} \mathcal{G}_{14,5,3}-\frac{1}{4} \mathcal{G}_{14,5,6}\nonumber \\
   &-\frac{1}{4} \mathcal{G}_{14,5,7}+\frac{1}{2}
   \mathcal{G}_{14,6,1}+\frac{1}{2} \mathcal{G}_{14,7,1}-\frac{1}{4} \mathcal{G}_{14,8,3}+\frac{1}{4} \mathcal{G}_{14,8,11}-\frac{1}{4}
   \mathcal{G}_{14,9,2}+\frac{1}{4} \mathcal{G}_{14,9,6}\nonumber \\
   &+\frac{1}{4} \mathcal{G}_{14,9,7}-\frac{1}{4} \mathcal{G}_{14,9,10}-\frac{1}{2}
   \mathcal{G}_{14,10,1}+\frac{1}{2} \mathcal{G}_{14,11,1}-\frac{1}{4} \mathcal{G}_{14,11,2}-\frac{1}{4}
   \mathcal{G}_{14,11,10}\nonumber \\
   &+\frac{1}{4} \mathcal{G}_{14,11,11}+\frac{1}{4} \mathcal{G}_{14,12,2}-\frac{1}{4}
   \mathcal{G}_{14,12,6}-\frac{1}{4} \mathcal{G}_{14,12,7}+\frac{1}{4} \mathcal{G}_{14,12,10}
\end{align}

\subsection{Top sector of family H}
In the following formula note that letters $l_i,\, i=\{7,8,9,10,16,17,18,19,25,26,27,28,29,30\}$ contain square roots in the \textit{underline kinematic variables} $S_{ij}$, as shown explicitly in \eqref{eq:alphabetH}.

\begin{align}
    g_{18,\,3}^H=&\frac{1}{4} L_2 \mathcal{G}_{27,3}-\frac{1}{4} L_4 \mathcal{G}_{27,3}-\frac{1}{4} L_5 \mathcal{G}_{27,3}-\frac{1}{4} L_3
   \mathcal{G}_{27,9}+\frac{1}{4} L_4 \mathcal{G}_{27,9}+\frac{1}{4} L_6 \mathcal{G}_{27,9}-\frac{1}{4} L_3 \mathcal{G}_{27,10}\nonumber \\
   &+\frac{1}{4} L_4
   \mathcal{G}_{27,10}+\frac{1}{4} L_6 \mathcal{G}_{27,10}-\frac{1}{4} L_2 \mathcal{G}_{27,11}+\frac{1}{4} L_4
   \mathcal{G}_{27,11}+\frac{1}{4} L_5 \mathcal{G}_{27,11}+\frac{1}{4} L_2 \mathcal{G}_{27,13}-\frac{1}{4} L_4
   \mathcal{G}_{27,13}\nonumber \\
   &-\frac{1}{4} L_5 \mathcal{G}_{27,13}-\frac{1}{4} L_2 \mathcal{G}_{27,18}+\frac{1}{4} L_3
   \mathcal{G}_{27,18}+\frac{1}{4} L_5 \mathcal{G}_{27,18}-\frac{1}{4} L_6 \mathcal{G}_{27,18}-\frac{1}{4} L_2
   \mathcal{G}_{27,19}+\frac{1}{4} L_3 \mathcal{G}_{27,19}\nonumber \\
   &+\frac{1}{4} L_5 \mathcal{G}_{27,19}-\frac{1}{4} L_6
   \mathcal{G}_{27,19}+\frac{1}{4} L_3 \mathcal{G}_{27,20}-\frac{1}{4} L_4 \mathcal{G}_{27,20}-\frac{1}{4} L_6
   \mathcal{G}_{27,20}+\frac{1}{4} L_2 \mathcal{G}_{27,24}-\frac{1}{4} L_3 \mathcal{G}_{27,24}\nonumber \\
   &-\frac{1}{4} L_5
   \mathcal{G}_{27,24}+\frac{1}{4} L_6 \mathcal{G}_{27,24}-\frac{1}{4} L_2 \mathcal{G}_{28,3}+\frac{1}{4} L_4 \mathcal{G}_{28,3}+\frac{1}{4}
   L_5 \mathcal{G}_{28,3}+\frac{1}{4} L_3 \mathcal{G}_{28,9}-\frac{1}{4} L_4 \mathcal{G}_{28,9}\nonumber \\
   &-\frac{1}{4} L_6 \mathcal{G}_{28,9}+\frac{1}{4}
   L_3 \mathcal{G}_{28,10}-\frac{1}{4} L_4 \mathcal{G}_{28,10}-\frac{1}{4} L_6 \mathcal{G}_{28,10}+\frac{1}{4} L_2
   \mathcal{G}_{28,11}-\frac{1}{4} L_4 \mathcal{G}_{28,11}-\frac{1}{4} L_5 \mathcal{G}_{28,11}\nonumber \\
   &-\frac{1}{4} L_2
   \mathcal{G}_{28,13}+\frac{1}{4} L_4 \mathcal{G}_{28,13}+\frac{1}{4} L_5 \mathcal{G}_{28,13}+\frac{1}{4} L_2
   \mathcal{G}_{28,18}-\frac{1}{4} L_3 \mathcal{G}_{28,18}-\frac{1}{4} L_5 \mathcal{G}_{28,18}+\frac{1}{4} L_6
   \mathcal{G}_{28,18}\nonumber \\
   &+\frac{1}{4} L_2 \mathcal{G}_{28,19}-\frac{1}{4} L_3 \mathcal{G}_{28,19}-\frac{1}{4} L_5
   \mathcal{G}_{28,19}+\frac{1}{4} L_6 \mathcal{G}_{28,19}-\frac{1}{4} L_3 \mathcal{G}_{28,20}+\frac{1}{4} L_4
   \mathcal{G}_{28,20}+\frac{1}{4} L_6 \mathcal{G}_{28,20}\nonumber \\
   &-\frac{1}{4} L_2 \mathcal{G}_{28,24}+\frac{1}{4} L_3
   \mathcal{G}_{28,24}+\frac{1}{4} L_5 \mathcal{G}_{28,24}-\frac{1}{4} L_6 \mathcal{G}_{28,24}+\frac{1}{4} L_5
   \mathcal{G}_{29,2}-\frac{1}{4} L_6 \mathcal{G}_{29,7}-\frac{1}{4} L_6 \mathcal{G}_{29,8}\nonumber \\
   &-\frac{1}{4} L_2 \mathcal{G}_{29,12}+\frac{1}{4} L_3
   \mathcal{G}_{29,12}+\frac{1}{4} L_5 \mathcal{G}_{29,14}-\frac{1}{4} L_5 \mathcal{G}_{29,15}+\frac{1}{4} L_2
   \mathcal{G}_{29,18}-\frac{1}{4} L_3 \mathcal{G}_{29,18}-\frac{1}{4} L_5 \mathcal{G}_{29,18}\nonumber \\
   &+\frac{1}{4} L_6
   \mathcal{G}_{29,18}+\frac{1}{4} L_2 \mathcal{G}_{29,19}-\frac{1}{4} L_3 \mathcal{G}_{29,19}-\frac{1}{4} L_5
   \mathcal{G}_{29,19}+\frac{1}{4} L_6 \mathcal{G}_{29,19}+\frac{1}{4} L_6 \mathcal{G}_{29,21}+\frac{1}{4} L_2
   \mathcal{G}_{29,22}\nonumber \\
   &-\frac{1}{4} L_3 \mathcal{G}_{29,22}-\frac{1}{4} L_2 \mathcal{G}_{29,24}+\frac{1}{4} L_3
   \mathcal{G}_{29,24}+\frac{1}{4} L_5 \mathcal{G}_{29,24}-\frac{1}{4} L_6 \mathcal{G}_{29,24}-\frac{1}{4} L_5
   \mathcal{G}_{30,2}+\frac{1}{4} L_6 \mathcal{G}_{30,7}\nonumber \\
   &+\frac{1}{4} L_6 \mathcal{G}_{30,8}+\frac{1}{4} L_2 \mathcal{G}_{30,12}-\frac{1}{4} L_3
   \mathcal{G}_{30,12}-\frac{1}{4} L_5 \mathcal{G}_{30,14}+\frac{1}{4} L_5 \mathcal{G}_{30,15}-\frac{1}{4} L_2
   \mathcal{G}_{30,18}+\frac{1}{4} L_3 \mathcal{G}_{30,18}\nonumber \\
   &+\frac{1}{4} L_5 \mathcal{G}_{30,18}-\frac{1}{4} L_6
   \mathcal{G}_{30,18}-\frac{1}{4} L_2 \mathcal{G}_{30,19}+\frac{1}{4} L_3 \mathcal{G}_{30,19}+\frac{1}{4} L_5
   \mathcal{G}_{30,19}-\frac{1}{4} L_6 \mathcal{G}_{30,19}-\frac{1}{4} L_6 \mathcal{G}_{30,21}\nonumber \\
   &-\frac{1}{4} L_2
   \mathcal{G}_{30,22}+\frac{1}{4} L_3 \mathcal{G}_{30,22}+\frac{1}{4} L_2 \mathcal{G}_{30,24}-\frac{1}{4} L_3
   \mathcal{G}_{30,24}-\frac{1}{4} L_5 \mathcal{G}_{30,24}+\frac{1}{4} L_6 \mathcal{G}_{30,24}+\frac{1}{4}
   \mathcal{G}_{27,3,2}\nonumber \\
   &-\frac{1}{4} \mathcal{G}_{27,3,4}-\frac{1}{4} \mathcal{G}_{27,3,5}+\frac{1}{4} \mathcal{G}_{27,3,14}+\frac{1}{4}
   \mathcal{G}_{27,9,4}+\frac{1}{4} \mathcal{G}_{27,9,5}-\frac{1}{4} \mathcal{G}_{27,9,7}-\frac{1}{4} \mathcal{G}_{27,9,8}\nonumber \\
   &+\frac{1}{4}
   \mathcal{G}_{27,10,4}+\frac{1}{4} \mathcal{G}_{27,10,5}-\frac{1}{4} \mathcal{G}_{27,10,7}-\frac{1}{4}
   \mathcal{G}_{27,10,8}-\frac{1}{4} \mathcal{G}_{27,11,15}+\frac{1}{4} \mathcal{G}_{27,13,2}-\frac{1}{4}
   \mathcal{G}_{27,13,4}\nonumber \\
   &-\frac{1}{4} \mathcal{G}_{27,13,5}+\frac{1}{4} \mathcal{G}_{27,13,14}-\frac{1}{4}
   \mathcal{G}_{27,16,2}+\frac{1}{4} \mathcal{G}_{27,16,4}+\frac{1}{4} \mathcal{G}_{27,16,5}-\frac{1}{4}
   \mathcal{G}_{27,16,14}+\frac{1}{4} \mathcal{G}_{27,16,15}\nonumber \\
   &-\frac{1}{4} \mathcal{G}_{27,17,2}+\frac{1}{4}
   \mathcal{G}_{27,17,4}+\frac{1}{4} \mathcal{G}_{27,17,5}-\frac{1}{4} \mathcal{G}_{27,17,14}+\frac{1}{4}
   \mathcal{G}_{27,17,15}-\frac{1}{4} \mathcal{G}_{27,18,2}+\frac{1}{4} \mathcal{G}_{27,18,7}\nonumber \\
   &+\frac{1}{4}
   \mathcal{G}_{27,18,8}-\frac{1}{4} \mathcal{G}_{27,18,14}-\frac{1}{4} \mathcal{G}_{27,19,2}+\frac{1}{4}
   \mathcal{G}_{27,19,7}+\frac{1}{4} \mathcal{G}_{27,19,8}-\frac{1}{4} \mathcal{G}_{27,19,14}+\frac{1}{4}
   \mathcal{G}_{27,20,21}\nonumber \\
   &+\frac{1}{4} \mathcal{G}_{27,24,15}-\frac{1}{4} \mathcal{G}_{27,24,21}-\frac{1}{4}
   \mathcal{G}_{27,25,4}-\frac{1}{4} \mathcal{G}_{27,25,5}+\frac{1}{4} \mathcal{G}_{27,25,7}+\frac{1}{4}
   \mathcal{G}_{27,25,8}-\frac{1}{4} \mathcal{G}_{27,25,21}\nonumber \\
   &-\frac{1}{4} \mathcal{G}_{27,26,4}-\frac{1}{4}
   \mathcal{G}_{27,26,5}+\frac{1}{4} \mathcal{G}_{27,26,7}+\frac{1}{4} \mathcal{G}_{27,26,8}-\frac{1}{4}
   \mathcal{G}_{27,26,21}+\frac{1}{4} \mathcal{G}_{27,29,2}-\frac{1}{4} \mathcal{G}_{27,29,7}\nonumber \\
   &-\frac{1}{4}
   \mathcal{G}_{27,29,8}+\frac{1}{4} \mathcal{G}_{27,29,14}-\frac{1}{4} \mathcal{G}_{27,29,15}+\frac{1}{4}
   \mathcal{G}_{27,29,21}+\frac{1}{4} \mathcal{G}_{27,30,2}-\frac{1}{4} \mathcal{G}_{27,30,7}-\frac{1}{4}
   \mathcal{G}_{27,30,8}\nonumber \\
   &+\frac{1}{4} \mathcal{G}_{27,30,14}-\frac{1}{4} \mathcal{G}_{27,30,15}+\frac{1}{4}
   \mathcal{G}_{27,30,21}-\frac{1}{4} \mathcal{G}_{28,3,2}+\frac{1}{4} \mathcal{G}_{28,3,4}+\frac{1}{4}
   \mathcal{G}_{28,3,5}-\frac{1}{4} \mathcal{G}_{28,3,14}\nonumber \\
   &-\frac{1}{4} \mathcal{G}_{28,9,4}-\frac{1}{4} \mathcal{G}_{28,9,5}+\frac{1}{4}
   \mathcal{G}_{28,9,7}+\frac{1}{4} \mathcal{G}_{28,9,8}-\frac{1}{4} \mathcal{G}_{28,10,4}-\frac{1}{4}
   \mathcal{G}_{28,10,5}+\frac{1}{4} \mathcal{G}_{28,10,7}\nonumber \\
   &+\frac{1}{4} \mathcal{G}_{28,10,8}+\frac{1}{4}
   \mathcal{G}_{28,11,15}-\frac{1}{4} \mathcal{G}_{28,13,2}+\frac{1}{4} \mathcal{G}_{28,13,4}+\frac{1}{4}
   \mathcal{G}_{28,13,5}-\frac{1}{4} \mathcal{G}_{28,13,14}+\frac{1}{4} \mathcal{G}_{28,16,2}\nonumber \\
   &-\frac{1}{4}
   \mathcal{G}_{28,16,4}-\frac{1}{4} \mathcal{G}_{28,16,5}+\frac{1}{4} \mathcal{G}_{28,16,14}-\frac{1}{4}
   \mathcal{G}_{28,16,15}+\frac{1}{4} \mathcal{G}_{28,17,2}-\frac{1}{4} \mathcal{G}_{28,17,4}-\frac{1}{4}
   \mathcal{G}_{28,17,5}\nonumber \\
   &+\frac{1}{4} \mathcal{G}_{28,17,14}-\frac{1}{4} \mathcal{G}_{28,17,15}+\frac{1}{4}
   \mathcal{G}_{28,18,2}-\frac{1}{4} \mathcal{G}_{28,18,7}-\frac{1}{4} \mathcal{G}_{28,18,8}+\frac{1}{4}
   \mathcal{G}_{28,18,14}+\frac{1}{4} \mathcal{G}_{28,19,2}\nonumber \\
   &-\frac{1}{4} \mathcal{G}_{28,19,7}-\frac{1}{4}
   \mathcal{G}_{28,19,8}+\frac{1}{4} \mathcal{G}_{28,19,14}-\frac{1}{4} \mathcal{G}_{28,20,21}-\frac{1}{4}
   \mathcal{G}_{28,24,15}+\frac{1}{4} \mathcal{G}_{28,24,21}+\frac{1}{4} \mathcal{G}_{28,25,4}\nonumber \\
   &+\frac{1}{4}
   \mathcal{G}_{28,25,5}-\frac{1}{4} \mathcal{G}_{28,25,7}-\frac{1}{4} \mathcal{G}_{28,25,8}+\frac{1}{4}
   \mathcal{G}_{28,25,21}+\frac{1}{4} \mathcal{G}_{28,26,4}+\frac{1}{4} \mathcal{G}_{28,26,5}-\frac{1}{4}
   \mathcal{G}_{28,26,7}\nonumber \\
   &-\frac{1}{4} \mathcal{G}_{28,26,8}+\frac{1}{4} \mathcal{G}_{28,26,21}-\frac{1}{4}
   \mathcal{G}_{28,29,2}+\frac{1}{4} \mathcal{G}_{28,29,7}+\frac{1}{4} \mathcal{G}_{28,29,8}-\frac{1}{4}
   \mathcal{G}_{28,29,14}+\frac{1}{4} \mathcal{G}_{28,29,15}\nonumber \\
   &-\frac{1}{4} \mathcal{G}_{28,29,21}-\frac{1}{4}
   \mathcal{G}_{28,30,2}+\frac{1}{4} \mathcal{G}_{28,30,7}+\frac{1}{4} \mathcal{G}_{28,30,8}-\frac{1}{4}
   \mathcal{G}_{28,30,14}+\frac{1}{4} \mathcal{G}_{28,30,15}-\frac{1}{4} \mathcal{G}_{28,30,21}\nonumber \\
   &-\frac{1}{4}
   \mathcal{G}_{29,2,4}-\frac{1}{4} \mathcal{G}_{29,2,5}+\frac{1}{4} \mathcal{G}_{29,6,2}+\frac{1}{4} \mathcal{G}_{29,6,14}-\frac{1}{4}
   \mathcal{G}_{29,6,15}+\frac{1}{4} \mathcal{G}_{29,7,4}+\frac{1}{4} \mathcal{G}_{29,7,5}\nonumber \\
   &+\frac{1}{4} \mathcal{G}_{29,8,4}+\frac{1}{4}
   \mathcal{G}_{29,8,5}-\frac{1}{4} \mathcal{G}_{29,12,2}+\frac{1}{4} \mathcal{G}_{29,12,7}+\frac{1}{4}
   \mathcal{G}_{29,12,8}-\frac{1}{4} \mathcal{G}_{29,12,14}-\frac{1}{4} \mathcal{G}_{29,14,4}\nonumber \\
   &-\frac{1}{4}
   \mathcal{G}_{29,14,5}-\frac{1}{4} \mathcal{G}_{29,16,2}+\frac{1}{4} \mathcal{G}_{29,16,4}+\frac{1}{4}
   \mathcal{G}_{29,16,5}-\frac{1}{4} \mathcal{G}_{29,16,14}+\frac{1}{4} \mathcal{G}_{29,16,15}-\frac{1}{4}
   \mathcal{G}_{29,17,2}\nonumber \\
   &+\frac{1}{4} \mathcal{G}_{29,17,4}+\frac{1}{4} \mathcal{G}_{29,17,5}-\frac{1}{4}
   \mathcal{G}_{29,17,14}+\frac{1}{4} \mathcal{G}_{29,17,15}+\frac{1}{4} \mathcal{G}_{29,18,2}-\frac{1}{4}
   \mathcal{G}_{29,18,7}-\frac{1}{4} \mathcal{G}_{29,18,8}\nonumber \\
   &+\frac{1}{4} \mathcal{G}_{29,18,14}+\frac{1}{4}
   \mathcal{G}_{29,19,2}-\frac{1}{4} \mathcal{G}_{29,19,7}-\frac{1}{4} \mathcal{G}_{29,19,8}+\frac{1}{4}
   \mathcal{G}_{29,19,14}+\frac{1}{4} \mathcal{G}_{29,22,15}-\frac{1}{4} \mathcal{G}_{29,22,21}\nonumber \\
   &-\frac{1}{4}
   \mathcal{G}_{29,23,7}-\frac{1}{4} \mathcal{G}_{29,23,8}+\frac{1}{4} \mathcal{G}_{29,23,21}-\frac{1}{4}
   \mathcal{G}_{29,24,15}+\frac{1}{4} \mathcal{G}_{29,24,21}-\frac{1}{4} \mathcal{G}_{29,25,4}-\frac{1}{4}
   \mathcal{G}_{29,25,5}\nonumber \\
   &+\frac{1}{4} \mathcal{G}_{29,25,7}+\frac{1}{4} \mathcal{G}_{29,25,8}-\frac{1}{4}
   \mathcal{G}_{29,25,21}-\frac{1}{4} \mathcal{G}_{29,26,4}-\frac{1}{4} \mathcal{G}_{29,26,5}+\frac{1}{4}
   \mathcal{G}_{29,26,7}+\frac{1}{4} \mathcal{G}_{29,26,8}\nonumber \\
   &-\frac{1}{4} \mathcal{G}_{29,26,21}+\frac{1}{4}
   \mathcal{G}_{30,2,4}+\frac{1}{4} \mathcal{G}_{30,2,5}-\frac{1}{4} \mathcal{G}_{30,6,2}-\frac{1}{4} \mathcal{G}_{30,6,14}+\frac{1}{4}
   \mathcal{G}_{30,6,15}-\frac{1}{4} \mathcal{G}_{30,7,4}\nonumber \\
   &-\frac{1}{4} \mathcal{G}_{30,7,5}-\frac{1}{4} \mathcal{G}_{30,8,4}-\frac{1}{4}
   \mathcal{G}_{30,8,5}+\frac{1}{4} \mathcal{G}_{30,12,2}-\frac{1}{4} \mathcal{G}_{30,12,7}-\frac{1}{4}
   \mathcal{G}_{30,12,8}+\frac{1}{4} \mathcal{G}_{30,12,14}\nonumber \\
   &+\frac{1}{4} \mathcal{G}_{30,14,4}+\frac{1}{4}
   \mathcal{G}_{30,14,5}+\frac{1}{4} \mathcal{G}_{30,16,2}-\frac{1}{4} \mathcal{G}_{30,16,4}-\frac{1}{4}
   \mathcal{G}_{30,16,5}+\frac{1}{4} \mathcal{G}_{30,16,14}-\frac{1}{4} \mathcal{G}_{30,16,15}\nonumber \\
   &+\frac{1}{4}
   \mathcal{G}_{30,17,2}-\frac{1}{4} \mathcal{G}_{30,17,4}-\frac{1}{4} \mathcal{G}_{30,17,5}+\frac{1}{4}
   \mathcal{G}_{30,17,14}-\frac{1}{4} \mathcal{G}_{30,17,15}-\frac{1}{4} \mathcal{G}_{30,18,2}+\frac{1}{4}
   \mathcal{G}_{30,18,7}\nonumber \\
   &+\frac{1}{4} \mathcal{G}_{30,18,8}-\frac{1}{4} \mathcal{G}_{30,18,14}-\frac{1}{4}
   \mathcal{G}_{30,19,2}+\frac{1}{4} \mathcal{G}_{30,19,7}+\frac{1}{4} \mathcal{G}_{30,19,8}-\frac{1}{4}
   \mathcal{G}_{30,19,14}-\frac{1}{4} \mathcal{G}_{30,22,15}\nonumber \\
   &+\frac{1}{4} \mathcal{G}_{30,22,21}+\frac{1}{4}
   \mathcal{G}_{30,23,7}+\frac{1}{4} \mathcal{G}_{30,23,8}-\frac{1}{4} \mathcal{G}_{30,23,21}+\frac{1}{4}
   \mathcal{G}_{30,24,15}-\frac{1}{4} \mathcal{G}_{30,24,21}+\frac{1}{4} \mathcal{G}_{30,25,4}\nonumber \\
   &+\frac{1}{4}
   \mathcal{G}_{30,25,5}-\frac{1}{4} \mathcal{G}_{30,25,7}-\frac{1}{4} \mathcal{G}_{30,25,8}+\frac{1}{4}
   \mathcal{G}_{30,25,21}+\frac{1}{4} \mathcal{G}_{30,26,4}+\frac{1}{4} \mathcal{G}_{30,26,5}-\frac{1}{4}
   \mathcal{G}_{30,26,7}\nonumber \\
   &-\frac{1}{4} \mathcal{G}_{30,26,8}+\frac{1}{4} \mathcal{G}_{30,26,21}
\end{align}


\end{document}